\documentclass[a4paper,11pt,reqno,twoside]{poincare}
\usepackage{amscd,amsmath,a4,amsfonts,amssymb,amsthm,enumerate,,mathdots}

\usepackage[table]{xcolor}

\usepackage{hyperref}
\usepackage[font=small]{caption}
\usepackage{subfigure} 
\usepackage{xcolor}
\usepackage{caption}
\usepackage{amsthm}
\usepackage{mathrsfs} 
\usepackage{enumerate}
\usepackage{multirow}
\usepackage{tabularx}
\usepackage{float}
\usepackage{color}
\usepackage{comment}

\renewcommand{\theequation}{\arabic{section}.\arabic{equation}}

\def\R{\mathbb{R}}

\def\G{\mathcal{G}_R}
\def\GO{\mathcal{G}}

\def\RR{\mathscr{R}}

\def\Diff{\mathcal{T}}

\def\RR{\mathscr{R}}

\newcommand{\TMP} {\mathbb T}

\newcommand{\inv}{\dagger}

\DeclareMathOperator*{\argmin}{argmin}

\usepackage{graphicx}
\graphicspath{{figures/}{./}}

\def\RR{\mathbb{R}}

\newcommand\blfootnote[1]{%
  \begingroup
  \renewcommand\thefootnote{}\footnote{#1}%
  \addtocounter{footnote}{-1}%
  \endgroup
}
\usepackage[hang,splitrule]{footmisc}



\theoremstyle{plain}
\newtheorem{theorem}{\bf Theorem}[section]
\newtheorem{lemma}[theorem]{\bf Lemma}
\theoremstyle{remark}
\newtheorem{definition}[theorem]{\bf Definition}

\newtheorem{example}[theorem]{\bf Example}

\newtheorem{remark}[theorem]{\bf Remark}
\newtheorem{proposition}[theorem]{\bf Proposition}
\newtheorem{corollary}[theorem]{\bf Corollary}
\numberwithin{equation}{section}

\usepackage{fancyhdr, graphicx}
\usepackage[dvips]{epsfig}

\usepackage[margin=1.5in]{geometry}
\usepackage{etoolbox}
\patchcmd{\section}{\scshape}{\bfseries}{}{}
\makeatletter
\renewcommand{\@secnumfont}{\bfseries}
\makeatother

\renewcommand\abstractname{\textbf{\emph{Abstract}}}

\title[\tiny RSCDT and its applications in Image Classification]{ The Radon Signed Cumulative Distribution Transform and its applications in classification of Signed Images}

\author[\tiny {\emph{L. G\lowercase{ong}, S. L\lowercase{i}, N.S. P\lowercase{athan},  M.S.E. R\lowercase{abbi}, G.K. R\lowercase{ohde}, A.H.M. R\lowercase{ubaiyat} \lowercase{and} S.T\lowercase{hareja}}}]{Le Gong}
\address[L. Gong]{D\lowercase{epartment of } M\lowercase{athematics}, Vanderbilt University}
\email{le.gong@vanderbilt.edu}

\author[]{Shiying Li}
\address[S. Li]{D\lowercase{epartment of }M\lowercase{athematics}, University of North Carolina at Chapel Hill}
\email{shiyl@unc.edu}

\author[]{Naqib Sad Pathan}
\address[N.S. Pathan]{D\lowercase{epartment of }E\lowercase{lectrical 
 and Computer }E\lowercase{ngineering}, University of Virginia }
\email{qpb3vt@virginia.edu}

\author[]{Mohammad Shifat-E-Rabbi}
\address[M.S.E. Rabbi]{D\lowercase{epartment of }E\lowercase{lectrical and }C\lowercase{omputer }E\lowercase{ngineering}, North South University}
\email{rabbi.mohammad@northsouth.edu}

\author[]{Gustavo K. Rohde}
\address[G.K. Rohde]{D\lowercase{epartment of }B\lowercase{omedical} E\lowercase{ngineering}, D\lowercase{epartment of }E\lowercase{lectrical and }C\lowercase{omputer }E\lowercase{ngineering}, University of Virginia} 
\email{gustavo.rohde@gmail.com}

\author[]{Abu Hasnat Mohammad Rubaiyat}
\address[A.H.M. Rubaiyat]{D\lowercase{epartment of }E\lowercase{lectrical and }C\lowercase{omputer }E\lowercase{ngineering}, University of Virginia}
\email{ar3fx@virginia.edu}

\author[]{Sumati Thareja}
\address[S. Thareja]{D\lowercase{epartment of }M\lowercase{athematics}, Vanderbilt University}
\email{sumatithareja5@gmail.com}

\subjclass{xxC15, xxA38}
\keywords{Signed Radon Cumulative Distribution Transform, Image Representation, Image Classification}

\begin{document}


\maketitle\thispagestyle{plain}



\begin{abstract}\fontsize{8}{10}\selectfont
Here we describe a new image representation technique based on the mathematics of transport and optimal transport. The method relies on the combination of the well-known Radon transform for images and a recent signal representation method called the Signed Cumulative Distribution Transform. The newly proposed method generalizes previous transport-related image representation methods to arbitrary functions (images), and thus can be used in more applications. We describe the new transform, and some of its mathematical properties and demonstrate its ability to partition image classes with real and simulated data. In comparison to existing transport transform methods, as well as deep learning-based classification methods, the new transform more accurately represents the information content of signed images, and thus can be used to obtain higher classification accuracies. The implementation of the proposed method in Python language is integrated as a part of the software package PyTransKit \cite{pytranskit}
\end{abstract}

\blfootnote{\emph{Communicated by}.  xxxxxx xxxxxx}
\blfootnote{$^\dag$\emph{Corresponding author}}

\fontsize{10}{12}\selectfont
\section{Introduction}

Finding useful mathematical formulas for represending signal and image data can be critical for solving important engineering and scientific problems. Fourier representation methods, for example, can dramatically simplify the solution of shift invariant linear systems of equations (e.g. convolutions), and thus are extensively used to filter sound and other types of signals, in optics and image processing (e.g. deconvolution) and other important problems. Localized sparse representations (e.g. short time Fourier transforms, Wavelets) have been extensively utilized in signal compression, denoising, filtering and other
applications given their ability to summarize important signal features using few parameters.

In the past few years, new signal and image representation methods (i.e. transforms) based on the mathematics of optimal mass transport have emerged. Unlike the aforementioned methods (Fourier and Wavelet representations), which represent functions as linear combinations of basis functions or frame vectors, the emerging transport-based techniques are nonlinear representation methods. They have been extensively used to render problems that are nonlinear, non convex, and difficult to solve in signal domain into linear, convex problems that have closed form solutions in transport transform domain. Examples include reconstructing images from few measurements \cite{kolouri2015,cattell2017reconstructing}, estimating signal parameters such as time delay and frequency \cite{rubaiyat2020parametric}, as well as classifying data where they have been shown to be especially effective in allowing for the formulation of closed form, efficient, and accurate solution of a wide variety of signal and image classification problems \cite{rubaiyat2022end, shifat2021radon, shifat2023invariance,  zhuang2022local}. 

To date, several types of transforms based on optimal transport have appeared in the literature, nearly all can be interpreted as a variation of the linear optimal transport (LOT) technique \cite{kolouri2016continuous, wang2013linear}, which is naturally applicable to data that can be interpreted as probability distributions. These include 1D signal transforms for non-negative distributional data \cite{park2017cumulative} and signed distributions \cite{aldroubi2021signed}, as well as transforms for representing non-negative and normalized images in two and higher dimensions \cite{kolouri2016radon}. However, transport-based representation techniques for signed images in two or higher dimensions are yet to be described, and thus the success of transport-based classification methods in applications that involve signed images \cite{Gaussianedge,microscopy,scalogram} has yet to be replicated.

Signed signals in two or higher dimensions are frequently encountered in science and engineering applications.  One example being magnetic resonance imaging (MRI) where voltage coils acquire a series of 1D signals which are then reconstructed into images using the Fourier transform method, yielding complex functions that have positive and negative values \cite{brown2014magnetic}.  Other examples include classification problems that employ scalograms derived from various types of Wavelet transforms \cite{scalogram}. Background subtraction is also a commonly performed pre-processing operation for microscopy and other types of optical imaging, typically yielding signed data \cite{microscopy}. Filtering techniques are also often employed to pre-process data prior to processing and classification. When signed filters are used (e.g. edge-type detection filters, convolutional neural networks), the resulting pre-processed images become signed \cite{Gaussianedge}. 

Our goal in this manuscript is to describe a transport-based method for representing 2D signed images and demonstrate its ability to solve classification problems involving this type of data. The new image representation is denoted the Radon Signed Cumulative Distribution Transform (RSCDT) and consists of combining the recently developed signed cumulative distribution transform (SCDT) \cite{aldroubi2021signed} together with the well-known Radon transform \cite{radon1917determination}. We describe ensuing mathematical properties of the representation, as well as a new Wasserstein-type metric for signed images which can be useful in the solution of pattern recognition problems. We then adapt the transport-based classification problem statement appearing in recent publications involving unsigned images \cite{shifat2021radon} to signed images and describe its solution. Computational results showing the performance of the classification method based on the newly developed RSCDT in comparison to existing methods, including deep learning, are then shown, followed by concluding remarks. For readability, proofs for the given theorems and statements are provided in the appendix.

\section{Preliminaries}
\label{prelim}
\subsection{Notation}
\label{prelim_a}

Throughout the manuscript, we deal with signals $s$ assuming these to be square integrable in their respective domains. That is, we assume that $\int_{\Omega_s} |s(x)|^2 dx < \infty$, where $\Omega_s\subseteq\R$ is the domain over which $s$ is defined. In addition, we at times make use of the common notation: $\| s \|^2 = <s, s> = \int_{\Omega_s} s(x)^* s(x) dx = \int_{\Omega_s} |s(x)|^2 dx $, where $<\cdot, \cdot>$ is the inner product. Signals are assumed to be real, so the complex conjugate $^*$ does not play a role.
We will apply the same notation for functions whose input argument is two dimensional, i.e. images. Let $\mathbf{x} \in \Omega_s \subseteq \R^2$. A 2D continuous function representing the continuous image is denoted $s(\mathbf{x}), \mathbf{x} \in \Omega_s$. Signals or images are denoted $s^{(k)}$ when the class information is available, where the superscript $(k)$ represents the class label.

\begin{table}[]
\centering
\caption{\normalsize{Description of symbols}}
\label{table:symbols}
\begin{tabular}{ll}
\hline
Symbols                & Description    \\ \hline
$s(x)~/~s(\mathbf{x})$ & Signal / image \\
$\Omega_s$& Domain of $s$\\
$\widetilde{s}(t,\theta)$ & Radon transform of $s$\\
$\widehat{s}(x)~/~\widehat{s}(t,\theta)$ & CDT / R-CDT transform of $s$\\
$\mathscr{R}(\cdot)~/~\mathscr{R}^{-1}(\cdot)$ & Forward / inverse Radon transform operation\\
$g(x)$ & Strictly increasing and differentiable function\\
$g^{\theta}(t)$ & Strictly increasing and differentiable \\& function, indexed by an angle $\theta$ \\
$s\circ g$&$s(g(x))$: composition of $s(x)$ with $g(x)$\\
$\widetilde{s}\circ g^\theta$&$\widetilde{s}(g^\theta(t),\theta)$: composition of $\widetilde{s}(t,\theta)$ with $g^{\theta}(t)$\\& along the $t$ dimension of $\widetilde{s}(t,\theta)$\\
$\GO$ &  Set of increasing diffeomorphisms $g(x)$ \\
$\G$ & Set of increasing diffeomorphisms $g^{\theta}(t)$ \\&parameterized by $\theta$ with $\theta \in [0,\pi]$\\
$\Diff$& Set of all possible increasing diffeomorphisms \\&from $\R$ to $\R$\\

\hline
\end{tabular}
\vspace{-1em}
\end{table}

Below we will also make use of one dimensional (1D) increasing diffeomorphisms (one to one mapping functions), which are denoted as $g(x)$ for signals and $g^\theta(t)$ when they need to be parameterized by an angle $\theta$. The set of all possible increasing diffeomorphisms from $\R$ to $\R$ will be denoted as $\Diff$. 
Finally, at times we also utilize the `$\circ$' operator to denote composition. 
A summary of the symbols and notation used can be found in Table~\ref{table:symbols}.

\subsection{The Cumulative Distribution Transform (CDT)}
\label{prelim_b}
The CDT \cite{park2017cumulative} is an invertible nonlinear 1D signal transform from the space of smooth probability densities to the space of diffeomorphisms. The CDT morphs a given input signal, defined as a probability density function (PDF), into another PDF in such a way that the Wasserstein distance between them is minimized. More formally, let $s(x),x\in\Omega_s$ and $r(x),x\in\Omega_r$ define a given signal and a reference signal, respectively, which we consider to be appropriately normalized such that $s>0,r>0$, and $\int_{\Omega_s} s(x)dx=\int_{\Omega_r} r(x)dx=1$. The forward CDT transform\footnote{We are using a slightly different definition of the CDT than in \cite{park2017cumulative}. The properties of the CDT outlined here hold in both definitions.} of $s(x)$ with respect to $r(x)$ is given by the strictly increasing function $\widehat{s}(x)$ that satisfies
\begin{align} \label{CDT_integral}
    \int_{-\infty}^{\widehat{s}(x)}s(u)du=\int_{-\infty}^{x}r(u)du 
\end{align}
As described in detail in \cite{park2017cumulative}, the CDT is a nonlinear and invertible operation, with the inverse being
\begin{align}
s(x)=\frac{d\widehat{s}^{-1}(x)}{dx}r\left(\widehat{s}^{-1}(x)\right), ~\mbox{and}~\widehat{s}^{-1}(\widehat{s}(x)) = x\nonumber.
\end{align}
Therefore, $\hat{s}$ can be seen as a representation of the original function $s$. Moreover, like the Fourier transform \cite{bracewell} for example, the CDT has a number of properties which will help us render signal and image classification problems easier to solve. Finally, note that throughout the manuscript we use the uniform distribution for the reference function $r,$ i.e. 
\begin{equation}
      r(x) = \begin{cases}
        1 , \quad x \in \Omega_r\\
        0,  \quad \text{otherwise}
    \end{cases}
\end{equation}

{\bf{Property II-B.1}}~{\it{(Composition)}}: Let $s(x)$ denote a normalized signal and let $\widehat{s}(x)$ be the CDT of $s(x)$. The CDT of $s_g=g's\circ g$ is given by 
\begin{align}
    \widehat{s}_g=g^{-1}\circ\widehat{s}
\end{align}
Here, $g\in\Diff$ is an invertible and differentiable function (diffeomorphism), $g^\prime=dg(x)/dx$, and `$\circ$' denotes the composition operator with $s\circ g=s(g(x))$. For a proof, see Appendix~A in supplementary materials.

\begin{figure}
    \centering
    \includegraphics[width=12cm]{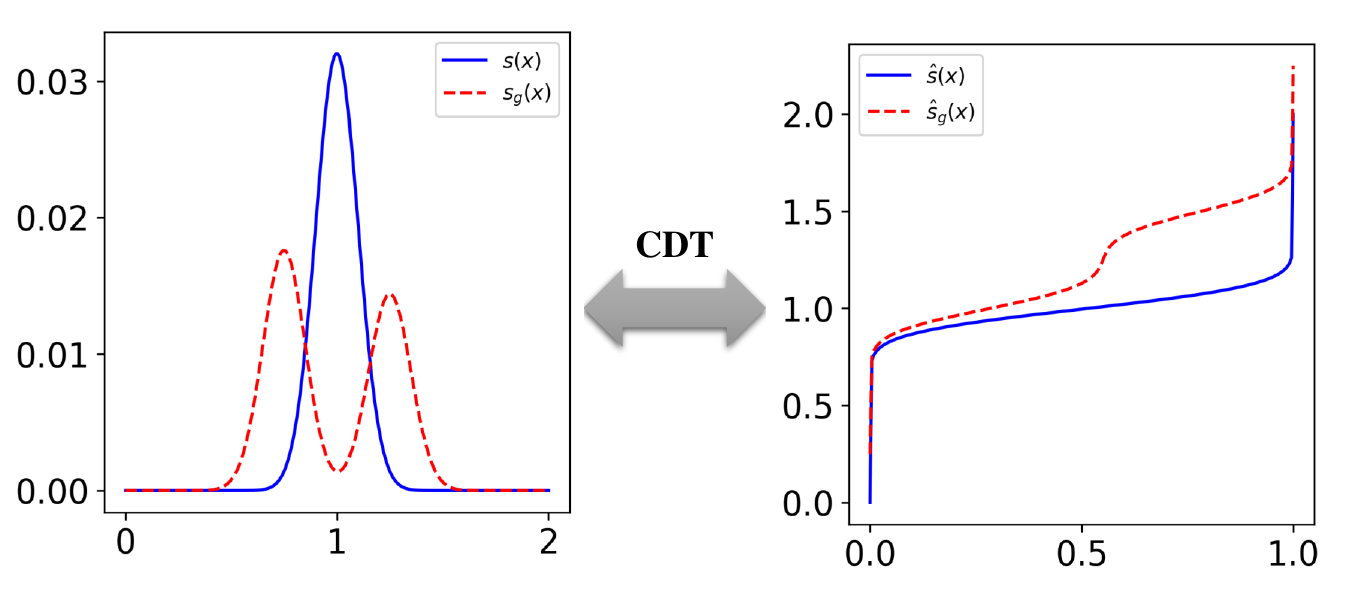}
    \caption{The cumulative distribution transform (CDT) of a signal (probability density function). Note that the CDT of an altered (transported) signal $s_g(x)$ (see text for definition) is related to the transform of $s$. In short, the CDT renders displacements into amplitude modulations in transform space.}
    \label{fig:cdt_comp}
    \vspace{-1em}
\end{figure}

The CDT composition property implies that, variations in a signal caused by applying $g(x)$ to the independent variable will change only the dependent variable in CDT space. This property is illustrated in Figure~\ref{fig:cdt_comp} where variations along both independent and dependent axis directions in original signal space become changes solely along the dependent axis in CDT space).\\\\
{\bf{Property II-B.2}} {\it{(Embedding)}}: 
The CDT induces an isometric embedding between the space of 1D signals  with the 2-Wasserstein metric and the space of their CDT transforms with a weighted-Euclidean metric \cite{park2017cumulative}, i.e., 
\begin{equation}\label{eq: cdtembedding}
	 W_2^2(s_1,s_2) =\left|\left|\left(\widehat{s}_1-\widehat{s}_2\right)\sqrt{r}\right|\right|_{L^2(\Omega_r)}^2,
\end{equation}
for all signals $s_1, s_2$. That is to say, if we wish to use the Wasserstein distance as a measure of similarity between $s_1,~s_2$, we can compute it as simply a weighted Euclidean norm in CDT space. 

The property above naturally links the CDT and Wasserstein distances for PDFs. Wasserstein \cite{villani2008optimal} distances are linked to optimal transport and have been used in a variety of applications in signal and image processing and machine learning (see \cite{kolouri2017optimal} for a recent review). 


\subsection{The Radon transform}
\label{prelim_c}
The Radon transform of an image $s(\mathbf{x}),\mathbf{x}\in\Omega_s\subset \R^2$, which we denote by $\widetilde{s}=\mathcal{R}(s)$, is defined as
\begin{eqnarray}
\widetilde{s}(t,\theta)&=&\int_{\Omega_s}s(\mathbf{x})\delta(t-\mathbf{x}\cdot \mathbf{\xi}_\theta)d\mathbf{x}
\end{eqnarray}
Here, $t$ is the perpendicular distance of a line from the origin and $\xi_\theta = [\cos(\theta),\sin(\theta)]^T$, where $\theta$ is the angle over which the projection is taken. 

Furthermore, using the Fourier Slice Theorem \cite{natterer2001mathematics, quinto2006introduction}, the inverse Radon transform $s=\mathcal{R}^{-1}(\widetilde{s})$ is defined as
\begin{eqnarray} \label{radon_inv}
s(\mathbf{x})&=&\int_0^\pi\int_{-\infty}^{\infty}\widetilde{s}(\mathbf{x}\cdot\xi_\theta-\tau,\theta)w(\tau)d\tau d\theta,
\end{eqnarray}
where $w$ is the ramp filter (i.e.,$(\mathscr Fw) (\xi) = |\xi|, \forall \xi$ ) and $\mathscr{F}$ is the Fourier transform.\\\\
{\bf{Property II-C.1}} ({\it{Intensity equality}}): Note that
\begin{eqnarray}
\int_{\Omega_s} s(\mathbf{x})d\mathbf{x} =\int_{-\infty}^{\infty}\widetilde{s}(t,\theta)dt, \;\;\;\;\; \forall\theta\in[0,\pi]
\label{eq:Rtheta}
\end{eqnarray}
which implies that $\int_{-\infty}^{\infty}\widetilde{s}(t,\theta_i)dt=\int_{-\infty}^{\infty}\widetilde{s}(t,\theta_j)dt$ for any two choices $\theta_i,\theta_j\in[0,\pi]$.

\subsection{Radon Cumulative Distribution Transform (R-CDT)}
\label{prelim_d}
The CDT framework was extended for 2D patterns (images as normalized density functions) through the sliced-Wasserstein distance in \cite{kolouri2016radon}, and was denoted as R-CDT. The main idea behind the R-CDT is to first obtain a family of one dimensional representations of a two dimensional probability measure (e.g., an image) through the Radon transform and then apply the CDT over the $t$ dimension in Radon transform space. More formally, let $s(\mathbf{x})$ and $r(\mathbf{x})$ define a given image and a reference image, respectively, which we consider to be appropriately normalized. The forward R-CDT of $s(\mathbf{x})$ with respect to $r(\mathbf{x})$ is given by the measure preserving function $\widehat{s}(t,\theta)$ that satisfies
\begin{align}\label{eq:rcdt}
    \int_{-\infty}^{\widehat{s}(t,\theta)}\widetilde{s}(u,\theta)du=\int_{-\infty}^{t}\widetilde{r}(u,\theta)du,~~~\forall\theta\in[0,\pi]
\end{align}

\begin{figure}
    \centering
    \includegraphics[width=14cm]{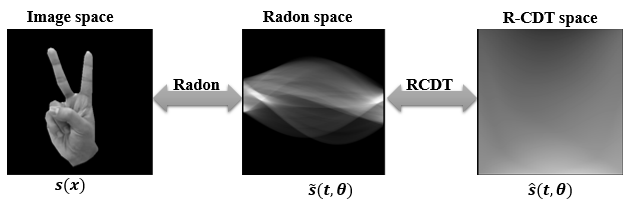}
    \caption{The process of calculating the Radon cumulative distribution transform (R-CDT) of an image $s(\mathbf{x})$ (defined as a 2-dimensional probability density function). The first step is to apply the Radon transform on $s(\mathbf{x})$ to obtain $\widetilde{s}(t,\theta)$. The R-CDT $\widehat{s}(t,\theta)$ is then obtained by applying the CDT over the $t$ dimension of $\widetilde{s}(t,\theta),~\forall\theta$.} 
    \label{fig:rcdt_process}
    \vspace{-1em}
\end{figure}

As in the case of the CDT, a transformed signal in R-CDT space can be recovered via the following inverse formula  \cite{kolouri2016radon}, 
\begin{align} \label{Radon_CDT_Inv}
s(\mathbf{x})=\RR^{-1}\left(\frac{\partial \widehat{s}^{-1}(t,\theta)}{\partial t}\widetilde{r}\left(\widehat{s}^{-1}(t,\theta),\theta\right)\right)
\end{align}
The process of calculating the R-CDT transform is shown in Figure~\ref{fig:rcdt_process}. As with the CDT, the R-CDT has a couple of properties outlined below which will be of interest when classifying images. Let us give some important definitions first. Let $G_R$ be the set defined as follows,
\begin{equation} \label{ConvexGroup}
\G = \{g = (g^{\theta})_{\theta \in [0,\pi]} : g^{\theta} : \R \rightarrow \R \text{ is a strictly increasing bijection } \forall \theta \in [0,\pi]\}
\end{equation}

\begin{definition}
   Let $\star : \G\times \G\rightarrow \G$ be an operator defined by $$(g\star h) (\cdot,\theta):= (g^{\theta}\circ h^{\theta})(\cdot) \quad \forall  \theta \in [0,\pi]$$.
\end{definition}

It is not hard to see that $(\G, \star)$ is a group (see \ref{append:GR_group}). 
\begin{definition}
   Let $H_{R}\subseteq \G$. We call $H_R$ a convex subgroup of $(\G, \star)$ if $H_R$ is a convex set and $(H_{R},\star)$ is a group.
\end{definition}

\noindent{\bf{Property II-D.1}} {\it{(Composition)}}: Let $s(\mathbf{x})$ denote an appropriately normalized image and let $\widetilde{s}(t,\theta)$ and $\widehat{s}(t,\theta)$ be the Radon transform and the R-CDT transform of $s(\mathbf{x})$, respectively. The R-CDT transform of $s_{g^\theta}=\mathscr{R}^{-1}\left(\left({g^\theta}\right)^\prime\widetilde{s}\circ {g^\theta}\right)$ is given by
\begin{align}
    \widehat{s}_{g^\theta}=({g^\theta})^{-1}\circ\widehat{s},
\end{align}
where $\left(g^\theta\right)^\prime=dg^\theta(t)/dt$, $\widetilde{s}\circ {g^\theta}:=\widetilde{s}(g^\theta(t),\theta)$, and $({g^\theta})^{-1}\circ\widehat{s}=({g^\theta})^{-1}(\widehat{s}(t,\theta))$.
Here for each fixed $\theta$, $g^\theta$ can be thought of an increasing and differentiable function with respect to $t$. The above equation hence follows from the composition property for 1D CDT. For a proof, see \cite{kolouri2016radon}.

The R-CDT composition property implies that, variations along both independent and dependent axis directions in an image, caused by applying $g^\theta(t)$ to the independent $t$ variable of its Radon transform, become changes solely along the dependent variable in R-CDT space.\\\\
{\bf{Property II-D.2}} {\it{(Embedding)}}:
R-CDT induces an isometric embedding between the space of images with  sliced-Wasserstein metric and the space of their R-CDT transforms with a weighted-Euclidean metric, i.e.,
\begin{equation}\label{eq: rcdtembedding}
	SW_2^2(s_1,s_2) = \left|\left|\left(\widehat{s}_1-\widehat{s}_2\right)\sqrt{\widetilde{r}}\right|\right|_{L^2(\Omega_{\widetilde r})}^2
\end{equation}
for all images $s_1$ and $s_2$. For a proof, see \cite{kolouri2016radon}. 

As the case with the 1D CDT shown above, the property above naturally links the R-CDT and sliced Wasserstein distances for PDFs and affords us a simple means of computing similarity among images \cite{kolouri2016radon}. We remark that throughout this manuscript we use the notation $\widehat s$ for both CDT or R-CDT transforms of a signal or image $s$ with respect to a fixed reference signal or image $r$, if a reference is not specified.



\subsection{Signed Cumulative Distribution Transform}
This section is a brief description of the SCDT (for details, see \cite{aldroubi2021signed}). Fix a reference signal $r$ that is a positive probability density. Let $s$ be a non-negative signal with $\|s\|_1=1.$ The cumulative distribution function $F_s$ of $s$ is defined as 
\begin{equation} \label {cumulation}
F_s(x) :=\int_{-\infty}^x s(t)dt.
\end{equation} 
The Cumulative Distribution Transform $\mathcal {C}(s)$ of $s$ with respect to $r$, as defined in eq. (\ref{CDT_integral}) (which implies $F_s(\widehat{s}(t)) = F_r(t)$)) \cite{park2017cumulative},  can also be written as (see \cite{aldroubi2021signed}) 
\begin{equation} \label{CDT_probs}
        \widehat{s}:={F_s}^{\inv}\circ {F_{r}}, 
    \end{equation}
where ${F_s}^{\inv}$ is the generalized inverse of $F_s$ defined by 
\begin{equation*}
        F^{\inv}(y) := \inf\{x: F(x)>y\}.
    \end{equation*}
Note that the generalized inverse comes in play here, because the CDF $F_s$ of $s$ need not be invertible.
Now, for a signed signal $s,$ the transform is defined by utilizing the Jordan decomposition of $s$, namely
\begin{align} \label{JordanDeco}
    s^+(t)=\max\{0,s(t)\} ,\hspace{0.1cm} s^-(t)=\max\{0,-s(t)\}
\end{align}

\noindent
Then, the SCDT of $s$ with respect to $r$ is defined as
    \begin{equation}\label{signed}
         \widehat{s} = \mathbb{T}_{r}(s) := \Big((s^{+})^\star, \|s^+\|_1, (s^{-})^\star, \|s^-\|_1\Big),
    \end{equation}
    where
    \begin{equation} \label{Normalized_CDT}
        (s^{\pm})^\star := \begin{cases}\widehat{\frac{s^\pm}{\|s^\pm\|_1}} & \text{ if }  s^{\pm} \text{ is non-trivial}\\
        0 & \text{ if } s^{\pm}=0 \, .
        \end{cases}  
    \end{equation}
    


Let $\mathcal{N} := \{f : \R \rightarrow \R \text{ non-decreasing }\} \times \R^+.$
Then, for a fixed reference $r,$
\begin{equation*}
    \mathbb{T}_r : L^{1}(\R) \rightarrow \mathcal{N} \times \mathcal{N}
\end{equation*} 
is a bijection, where for $(f,u,g,v) \in \mathcal{N} \times \mathcal{N}$ the inverse transform is given as,
\begin{equation} \label{SCDT_inv}
    (f,u,g,v) \longrightarrow \frac{u}{\|r\|_1} r(f^{\dagger}(\cdot)) (f^{\dagger})'(\cdot) - \frac{v}{\|r\|_1} r(g^{\dagger}(\cdot)) (g^{\dagger})'(\cdot)
\end{equation}
provided $f^{\dagger}$ and $g^{\dagger}$ are differentiatble.


 Note that, the SCDT expands on the concept of the CDT by accommodating functions with both positive and negative values. By splitting a signed signal into its positive and negative components, the SCDT allows for the application of the CDT independently to each part. This enables the transformation of signed signals while preserving important properties and characteristics. The introduction of the SCDT also introduces the concept of the generalized inverse. The generalized inverse behaves similarly to the regular inverse for invertible functions, however, it extends to situations where functions may not be invertible. This becomes necessary in the case of the SCDT, given that the Jordan decomposition used above (see \eqref{JordanDeco}) yields two non-negative signals $s^+$ and $s^-$ which might contain zeroes, and hence their cumulation (an analogue of CDF, for functions that are not probability distributions (see \cite{aldroubi2021signed})) stated in equation \eqref{cumulation} might not be stricly increasing and thus, might not admit an inverse. Therefore, the generalized inverse, is valuable in situations where a direct inverse may not exist, still allowing for the recovery and analysis of signals even when invertibility is not guaranteed.

By incorporating the SCDT and the notion of the generalized inverse, the framework becomes more versatile and applicable to a broader range of signals. Lets now see some of the important properties of SCDT, that are a more generalized version of the properties of the CDT.


\begin{lemma} (Composition Property) \label{Composition_Signed} \cite{aldroubi2021signed}
     Let $s \in L^{1}(\R)$, and let $g : \R \rightarrow \R$ be a strictly increasing surjection. Consider $s_g$ given by 
    $s_g(t) = g'(t) (s(g(t))$. Then,
     the SCDT of $s_g$ satisfies 
    $$\widehat{s}_g = \Big(g^{-1}\circ (s^{+})^{\star}, \| s^+\|_1, g^{-1}\circ (s^{-})^{\star}, \|s^-\|_1\Big).$$
\end{lemma}

In order to state the second lemma, we require a metric on the native space. This metric is a generalization of Wasserstein distance \cite{villani2003topics} defined for probability densities, to non-normalized signed signals.

\begin{definition} (Signed Wasserstein Metric)\label{distance}
Let $r, s\in L^1(\R)$ such that $\int r(t)  |t|^2 dt<\infty$ and $\int s(t)  |t|^2  dt<\infty$ , then

    \begin{equation} \label{SCDT_Metric}
        D^2_S(r,s) := D^{2}_{W^2}\left(r^+,s^+\right) + D^{2}_{W^2}\left({r}^-,s^-\right) 
    \end{equation}
    where $D^{2}_{W^2}\left(p,q\right) = d^{2}_{W^2}\left(\frac{p}{\|{p}\|_1},\frac{q}{\|q\|_1}\right) + \left(\|{p}\|_1 -  \|q\|_1\right)^2$ and $d^{2}_{W^2}(\cdot,\cdot)$ is the usual Wasserstein distance \cite{villani2003topics}.
\end{definition}

Utilizing the SCDT and the metric $D_S$, the following result is a generalization of a well-known isometry.

\begin{lemma} (Isometry) \label{isometry} \cite{aldroubi2021signed}
    Let $s_1,s_2\in L^1(\R)$ such that $\int s_1(t) |t|^2 dt<\infty$ and $\int s_2(t) |t|^2  dt<\infty$. Then, 
    \begin{align*}\label{wass_prob2}
         D^{2}_{S}(s_1,s_2) &= \|\widehat{s_1} -\widehat{s_2} \|^{2}_{(L^2(r)\times \R)^2}  \\
         &:= \|({s_1^+})^\star -({s_2^+})^\star \|^{2}_{L^2(r)} + \left( \|s_1^+\|_1 -\|s_2^+\|_1 \right)^2 \\
         & \; + \|({s_1^-})^\star -({s_2^-})^\star \|^{2}_{L^2(r)} + \left(\|s_1^-\|_1 -\|s_2^-\|_1 \right)^2,
    \end{align*}
where $\|\cdot\|_{L^2(r)}$ is the norm defined by
$$\|f\|_{L^2(r)}^2:=\int |f(x)|^2 \, r(x) \, dx .$$
\end{lemma}

Now that we have all the preliminaries in place, we define the Radon Signed Cumulative Distribution Transform and see some of its interesting properties that will later help us in classification of signed signals in 2D.


\begin{section} {Radon Signed Cumulative Distribution Transform}
\begin{figure}
    \centering
    \includegraphics[width=14 cm]{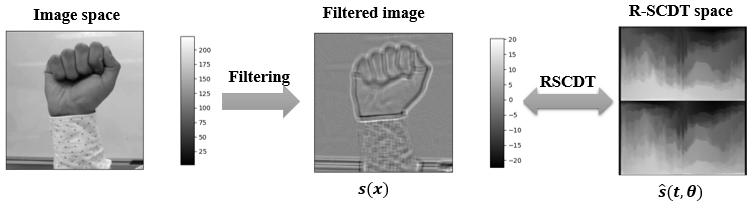}
    \caption{The process of obtaining the Radon Signed Cumulative Distribution Transform (R-SCDT) of a grayscale image with background is depicted. In the first step, a filtering technique is applied to suppress the background as well as to enhance the features of interest in the image. The filtered image contains negative pixel values as shown in the figure. To calculate the R-SCDT ($\widehat{s}(t,\theta)$) of the filtered image, a two-step process is followed. First, the Radon transform is applied to the filtered image which computes the line integrals at different angles. Next, the signed cumulative distribution transform (SCDT) is applied to each projection obtained from the Radon transform.} 
    \label{fig:rscdt_process}
    \vspace{-1em}
\end{figure}

Often many image pre-processing steps like background subtraction leave images having both positive and negative values. An example of such image is shown in Figure~\ref{fig:rscdt_process}. Here we expand the definition of the R-CDT provided above (see section \ref{prelim_d}) to allow us to deal with signed images. The idea can be simply summarized as the application of the SCDT to the Radon transform of a signed function. We describe below the mathematical properties of this new representation and show how they can be used to obtain classifiers which are at the same time accurate, closed form, and cheap to compute.

\begin{definition} (RSCDT) Let $r : \R^2 \rightarrow \R$ be the positive reference image and $s : \R^2 \rightarrow \R$ be the signed image, then Radon Signed CDT of $s$ with respect to $r$ is a composition of the Radon transform and SCDT for each projection angle. Formally, we first take the Radon transform of both the reference and the signal as,
\begin{equation*}
    \widetilde{r}(t,\theta) = \mathcal{R}(r(\textbf{x})) \quad \text{ and } \quad \widetilde{s}(t,\theta) = \mathcal{R}(s(\textbf{x})) .
\end{equation*}
Finally, the RSCDT of $s$ with respect to the reference $r,$ for each $\theta \in [0,\pi]$ is given as,
\begin{equation}
    \widehat{s}(\cdot, \theta) =  \mathbb{T}_{\tilde{r}(\cdot,\theta)}(\tilde{s}(\cdot,\theta)) = \Big(({\widetilde{s}}^+)^\star(\cdot, \theta), \|{\widetilde{s}}^+(\cdot,\theta)\|_1, ({\widetilde{s}}^-)^\star(\cdot, \theta), \|{\widetilde{s}}^-(\cdot, \theta)\|_1\Big)
\end{equation}
where $(\cdot)^\star$ and $ \mathbb{T}_{\tilde{r}(\cdot,\theta)}$ are as defined in \eqref{Normalized_CDT} and \eqref{SCDT_inv}, respectively.


    
\end{definition}

As in the case of its family transforms, a transformed image in the RSCDT space can be recovered via following inverse formula

\begin{equation*}
    s = \mathcal{R}^{-1} ([\TMP^{-1}_{\widetilde{r}(\cdot,\theta)}(\widehat{s}(\cdot,\theta))]_{\theta \in [0,\pi]})
\end{equation*}
where $\mathcal{R}^{-1}$ is the inverse of the Radon transform (see \eqref{radon_inv}) and $\TMP^{-1}$ is the inverse of the SCDT (see \eqref{SCDT_inv}).

\subsection{Properties of RSCDT}

Like its family transforms, RSCDT has a few properties outlined below which will be of interest when classifying images. 

\begin{proposition} (Composition Property) \label{composition}
Let $s(\bf{x})$ denote a signed image and let $\tilde{s}(t, \theta)$ and $\widehat{s}(t,\theta)$ be the Radon transform and RSCDT of $s(\bf{x})$ respectively. For $g\in \G$ and for each $\theta \in [0,\pi],$ define
\begin{equation}
    s_{g} = \mathcal{R}^{-1} ((g^{\theta})' \widetilde{s} \circ g^{\theta})
\end{equation}
\noindent where $(g^{\theta})'(t) = d{g^{\theta}}(t)/dt, \ \tilde{s} \circ g^{\theta}(\cdot) := \tilde{s}(g^{\theta}(\cdot),\theta).$ 
Then the RSCDT of $s_{g}$ is given by 
\begin{equation}
    \widehat{s}_g(t, \theta) = \Big((g^{\theta})^{-1} \circ (\widetilde{s}^+)^\star(t, \theta), \|{\widetilde{s}}^+(\cdot,\theta)\|_1, (g^{\theta})^{-1} \circ  ({\widetilde{s}}^-)^\star(t, \theta), \|\widetilde{s}^-(\cdot, \theta)\|_1\Big)
\end{equation}
\end{proposition}

Like in the previous versions of the transform, the composition property of the RSCDT implies that variations along both the independent and dependent axis directions in an image, caused by $g^{\theta}(t)$ to the independent variable in the Radon space becomes solely along the dependant variable in the RSCDT space. The following corollary, gives an explicit example of the composition property for an easy to understand variation of translation in the independent variables.

\begin{corollary}
Let $s(\bf{x})$ denote a signed image and let $\tilde{s}(t, \theta)$ and $\widehat{s}(t,\theta)$ be the Radon transform and RSCDT of $s(\bf{x})$ respectively. Let $$s_1(x_1,x_2) = s(x_1 - x_0, y_1-y_0),$$
Then the RSCDT $\widehat{s_1}(\cdot, \theta)$ of $s_1$ for every $\theta \in [0,\pi]$ is given by 

\begin{equation}
    \widehat{s_1}(t, \theta) = \Big((g^{\theta})^{-1} \circ (\widetilde{s}^+)^\star(t, \theta), \|\widetilde{s}^+(\cdot,\theta)\|_1, (g^{\theta})^{-1} \circ  (\widetilde{s}^-)^\star(t, \theta), \|\widetilde{s}^-(\cdot, \theta)\|_1.\Big)
\end{equation}
 where, $g^{\theta}(t) = t - x_0\cos\theta - y_0\sin\theta.$
\end{corollary}

In certain applications data sets can be rendered convex by family transforms of RSCDT in the respective transform domains. For example, we saw in case of 1D in \cite{park2017cumulative} sets generated by translations of a template signal can have a complex geometry in the signal domain. However, they have a very simple convex structure in the transform domain. This convexification property is useful in classification problems since two disjoint convex data sets can be separated by a linear classifier. The following convexity property is a generalization of the convexity property that was proved in \cite{aldroubi2021signed} for signed signals in 1D. 

\begin{proposition} (Convexification Property) \label{convexification}
Let $\phi(\bf{x})$ denote a signed image and let $\tilde{\phi}(t, \theta)$ and $\widehat{\phi}(t,\theta)$ be the Radon transform and RSCDT of $\phi(\bf{x})$ respectively. Let $H_R\subseteq \G$. Consider the generative model,
\begin{equation} \label{gen_model}
    \mathbb{S}_{\phi,H_R} := \{ \phi_g : \phi_g = \mathcal{R}^{-1} ((g^{\theta})' \widetilde{\phi} \circ g^{\theta})  ,  \  g \in H_R\}
\end{equation}
If $H_R^{-1}$ is a convex set then $\widehat{\mathbb{S}}_{\phi,H_R} := \{\widehat{\phi}_g : \phi_g  \in \mathbb{S}_{\phi,H_R}\}$ is convex.

\end{proposition}

Next is a corollary to the convexification property above that is based on the fact that if $G$ is a group then $G = G^{-1}.$

\begin{corollary}
    Let $\phi(\bf{x})$ be as in the above proposition. Let $H_R\subseteq \G$ such that $H_R$ is a convex group. 
If $H_R$ is convex, then $\widehat{\mathbb{S}}_{\phi,H_R} := \{\widehat{\phi}_g : \phi_g  \in \mathbb{S}_{\phi,H_R}\}$ (see \eqref{gen_model}) is convex.
\end{corollary}

\subsection{Metric Structure} The Wasserstein distance in the space $\mathcal{P}(\R)$ is intimately related to the $L^2$ distance in the transport transform domain as shown in \cite{aldroubi2020partitioning, thorpeNotes, villani2008optimal}. We then saw a Wasserstein like metric defined in \cite{aldroubi2021signed} for signed signals and its relationship with $L^2$ distance in the transform domain for 1D signed signals. This relation has proven to be useful in some applications since it render certain optimization problems involving the Wasserstein like distances into standard least squares optimization problems. In this section, we generalize the Wasserstein like distance defined in \cite{aldroubi2021signed} for signed measures and the sliced Wasserstein distance \cite{kolouri2016radon} to define a metric space structure on the space of signed signals (images) in 2D. We then see an analogue of the isometry \cite{aldroubi2021signed} between the metric on signed images and an analogue of the $L^2$ distance in the transform domain.

\begin{definition} (Signed Sliced-Wasserstein Distance) \label{signed_sliced_wasser}
For $s_1,s_2 : \R^2 \rightarrow \R$ the distance is defined as
\begin{align*}
    D(s_1,s_2) :&= \Big( \int_{0}^{\pi}  
    D_{S}^2 \left({\widetilde{s_{1}}}(\cdot,\theta),{\widetilde{s_{2}}}(\cdot,\theta)\right) d\theta  \Big)^{\frac{1}{2}}
    \\ &= \Bigg( \int_{0}^{\pi}  
    D_{W^2}^2 \left({\widetilde{s_{1}}}^+(\cdot,\theta),{\widetilde{s_{2}}}^+(\cdot,\theta)\right) d\theta + \int_{0}^{\pi} D_{W^2}^2 \left({\widetilde{s_{1}}}^-(\cdot,\theta),{\widetilde{s_{2}}}^-(\cdot,\theta)\right)
     d\theta \Bigg)^{\frac{1}{2}},
\end{align*}
where $D_S$ and $D_{W^2}$ are as defined in \eqref{SCDT_Metric}.


\end{definition}

\begin{proposition} \label{isometry_rscdt}
(Isometry/Embedding) RSCDT induces an isometric embedding between the space of images with Signed Sliced-Wasserstein metric defined above and the space of their RSCDT transforms with a Euclidean-type metric,
\begin{equation} \label{eq:rscdt_iso1}
    D(s_1,s_2) = \Bigg( \int_{\theta = 0}^{\pi}  \|\widehat{s_1}(\cdot,\theta) - \widehat{s_2}(\cdot,\theta)\|^{2}_{(L^2{(\widetilde{r}(\cdot,\theta))} \times \R)^2} \, d\theta \Bigg)^{\frac{1}{2}},
\end{equation}
A compact notation for the above equation would be,
\begin{equation}
    D(s_1,s_2) = \|\widehat{s_1} - \widehat{s_2}\|_{(L^2{(\widetilde{r})} \times L^2[0,\pi])^2}
\end{equation} where 
\begin{align*}
    \|\widehat{s}\|^{2}_{(L^2{(\widetilde{r})} \times L^2[0,\pi])^2} &:=
    \int_{0}^{\pi} \| \widehat{s}(\cdot,\theta)\|^{2}_{(L^2{(\widetilde{r}(\cdot, \theta))} \times \R)^2} \, d\theta
     \\ &= \int_{0}^{\pi} \Bigg(\int_{\R} |\widetilde{s}^+(t,\theta)|^2 \, \widetilde{r}(t,\theta) \, dt + \|\widetilde{s}^{+}(\cdot,\theta)\|_{1}^2 \\ &+ \int_{\R} |\widetilde{s}^-(t,\theta)|^2 \, \widetilde{r}(t,\theta) \, dt + \|\widetilde{s}^{-}(\cdot,\theta)\|_{1}^2 \Bigg) \, d\theta
\end{align*}


\end{proposition}
    
\end{section}

\section{Image Classification using RSCDT}
Here we discuss a mathematical model-based problem statement for the type of classification problems we discuss in this paper. Furthermore, we illustrate how the composition and convexification properties of the RSCDT play a crucial role in facilitating the classification of images.

\subsection{Signal Class Model and Problem Statement}
\label{plm_state}
In numerous applications, we focus on classifying images that are instances of a certain prototype (or template) observed under some often unknown deformation patterns.
Let's consider the problem of classifying handwritten digits, such as the MNIST dataset \cite{lecun1998gradient}. In such datasets, it is reasonable to assume that each observed digit image can be considered as an instance of a template (or templates) subjected to unknown deformations or variations. For instance, a suitable model for a particular class in the dataset, like the digit 1, would involve a fixed pattern for the digit 1 but with different translations, meaning the digit can appear randomly positioned within the image's field of view. Alternatively, the digit could also exhibit variations in size or slight nonrigid deformations. The following mathematical model for image (2D signals) classes formalize these concepts.

\begin{definition}[2D signal class model]
\label{def:gen_model_2d}
    Let $\G\subset\Diff$ be the set of confounds. The 2D mass (image intensity) preserving class model for the $k^{\mbox{th}}$ class is defined to be the set 
\begin{eqnarray}
\mathbb{S}^{(k)}=\left\{s_j^{(k)}|s_j^{(k)}=\mathcal{R}^{-1}\left(\left({g_j^\theta}\right)^\prime\widetilde{\varphi}^{(k)}\circ g^\theta_j\right), \forall g_j\in\G \right\},
\label{eq:2dgenerative_model}
\end{eqnarray}
\end{definition}
\noindent where $\varphi^{(k)}$ is the template pattern of class $k$ and $s_j^{(k)}$ is the $j$-th image from that class. Given the signal class model, the mathematical description of the classification problem is defined as:
\begin{definition}[Classification problem]
\label{def:classification}
Let $\G\subset\Diff$ denotes the set of confounds and $\mathbb{S}^{(k)}$ be defined as in equation~\eqref{eq:2dgenerative_model}. Given training samples $\{s^{(1)}_1, s^{(1)}_2, \cdots\}$ (class 1), $\{s^{(2)}_1, s^{(2)}_2, \cdots\}$ (class 2), $\cdots$ as training data, determine the class $(k)$ of an unknown image $s$.
\end{definition}

\subsection{Proposed Solution}
\label{soln}
The signal class model specified in equation (\ref{eq:2dgenerative_model}) typically produces signal classes that are nonconvex. However, the RSCDT can simplify the data geometry and thereby simplify the classification problem. Hence, following the methodology proposed in \cite{shifat2021radon}, we employ the composition property of the RSCDT on $s_j^{(k)}$ in eq. (\ref{eq:2dgenerative_model}), which yields the signal class model in the transform domain as:
\begin{eqnarray}
\widehat{\mathbb{S}}^{(k)}&=&\{\widehat{s}_j^{(k)}|\widehat{s}_j^{(k)}={\left(g_j^\theta\right)}^{-1}\circ \widehat{\varphi}^{(k)}, \forall g_j\in\G \}. 
\label{eq:2dgenerative_model_rscdt}
\end{eqnarray}
By applying the convexification property of the RSCDT, as described in Proposition \ref{convexification}, it is evident that the class model given in eq. (\ref{eq:2dgenerative_model_rscdt}) forms a convex set if $\G$ is a convex group \cite{shifat2021radon}. Furthermore, since the RSCDT is a one-to-one mapping, it follows that if $\mathbb{S}^{(k)} \cap \mathbb{S}^{(p)}=\varnothing$, then $\widehat{\mathbb{S}}^{(k)} \cap  \widehat{\mathbb{S}}^{(p)}=\varnothing$. Consequently, we define a subspace generated by the convex set $\widehat{\mathbb{S}}^{(k)}$ as follows:
\begin{equation}
    \widehat{\mathbb{V}}^{(k)} = \text{span}\left(\widehat{\mathbb{S}}^{(k)}\right) = \left\{\sum_{j\in \mathbf{J}}\alpha_j\widehat{s}_j^{(k)}|\alpha_j\in\mathbb{R},\mathbf{J} \text{ is finite} \right\}.
    \label{eq:subspace}
\end{equation}
It can be demonstrated, under specific assumptions, that the convex space associated with a particular class in the transformed domain does not overlap with the subspace corresponding to a different class \cite{shifat2021radon}, i.e. $\widehat{\mathbb{S}}^{(k)}\cap\widehat{\mathbb{V}}^{(p)}=\varnothing,k\neq p$.
It follows from the analysis that for a test sample $s$ generated according to the mathematical model for an unknown class $k$, we have $d^2( \widehat{s},\widehat{\mathbb{V}}^{(k)}) \sim 0$, and $d^2( \widehat{s},\widehat{\mathbb{V}}^{(p)})>0$ for $p\neq k$. In this context, $d^2(\cdot,\cdot)$ represents the Euclidean distance between $\widehat{s}$ and the nearest point in $\widehat{\mathbb{S}}^{(k)}$ or $\widehat{\mathbb{V}}^{(k)}$. Therefore, the unknown class of the test image can be predicted by employing a nearest subspace search method in the RSCDT space:
\begin{equation}
    \argmin_k~d^2(\widehat{s}, \widehat{\mathbb{V}}^{(k)}).
    \label{eq:soln1}
\end{equation}

\subsubsection{Training Algorithm}
Based on the aforementioned analysis, we propose a non-iterative training algorithm following the approach outlined in \cite{shifat2021radon}. The algorithm proceeds as follows: First, the transforms of the training samples are computed to obtain $\widehat{\mathbb{S}}^{(k)}$ for all classes. Next, we approximate $\widehat{\mathbb{V}}^{(k)}$ by taking the span of $\widehat{\mathbb{S}}^{(k)}$ resulting in, $\widehat{\mathbb{V}}^{(k)} =\text{span}\left\{\widehat{s}_1^{(k)}, \widehat{s}_2^{(k)}, ... \right\}$. The set $\left\{\widehat{s}_1^{(k)}, \widehat{s}_2^{(k)}, ... \right\}$ is orthogonalized to obtain the set of basis vectors $\{b_1^{(k)}, b_2^{(k)}, ... \}$ that spans the space $\widehat{\mathbb{V}}^{(k)}$ for class $k$. Finally, the matrix $B^{(k)}$ is formed with the computed basis vectors as its columns, i.e., $B^{(k)} = \left[b_1^{(k)}, b_2^{(k)}, ... \right]$. This matrix is later used for predicting the class label of unknown samples. It is important to note that while this algorithm is designed for classifying signal patterns that can be modeled as instances of a particular template (as defined in equation \ref{eq:2dgenerative_model}), the proposed RSCDT-NS method does not require the estimation of the template $\varphi^{(k)}$ for each class. Instead, one can directly use the training samples for a specific class $k$ to generate the matrix $B^{(k)}$ using the aforementioned algorithm. In addition, it is important to note that the convexity results, and thus our ability to model signed image classes as vector spaces, is independent of the reference chosen to define the RSCDT.

\subsubsection{Testing Algorithm}
Let's consider the problem of predicting the class of a test image $s$ using the proposed RSCDT-NS method. Firstly, we obtain the RSCDT $\widehat{s}$ of the test sample $s$. Next, we estimate the distance between $\widehat{s}$ and the subspace model for each class using $d^{2}(\widehat{s}, \widehat{\mathbb{V}}^{(k)})\sim \| \widehat{s} - B^{(k)}{B^{(k)}}^T\widehat{s} \|_{L_2}^2$. The class of $\widehat{s}$ is then estimated by solving the optimization problem: 
\begin{equation}
    \argmin_k~\|\widehat{s} - A^{(k)}\widehat{s}\|_{L_2}^2,
    \label{eq:test_alg}
\end{equation} 
where $A^{(k)} = B^{(k)}{B^{(k)}}^T$ represents an orthogonal projection matrix onto the subspace spanned by the columns of $B^{(k)}$.

\section{Computational Results}\medskip
We evaluated our proposed image classification method on three datasets, showcasing its effectiveness. In the first dataset, a simulated dataset comprising two classes, we demonstrate a scenario where the existing transport-based classification method (e.g., RCDT-NS) fails to differentiate between the two classes. However, our proposed method (RSCDT-NS) achieves 100\% accuracy in classifying the two classes. Next, we present another scenario in which negative pixel values can arise during image classification, highlighting the need for a signed transport-based method to handle such situations effectively. In the third case, we applied our proposed method to a real dataset and observed that the signed transport-based method, RSCDT-NS, outperforms both the unsigned methods and state-of-the-art convolutional neural network-based algorithms. 

\subsection{Result on simulated data}
\begin{figure}[htbp]
    {\includegraphics[scale=0.7]{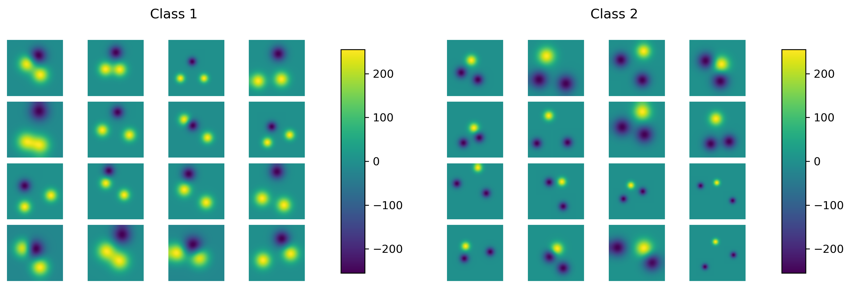}}

	\caption{(a) Simulated images from class 01, showcasing three circular shapes with randomized sizes and positions. These circular shapes are distinguished by two positive pixel values and one negative pixel value. (b) Simulated images from class 02, featuring three circular shapes with randomized sizes and positions. Unlike class 01, two of these circular shapes exhibit negative pixel values, while the remaining one displays positive pixel values.}
	\label{fig:Simulated}
\end{figure}

This study employs simulated data consisting of two categories of image data. The first category comprises three circular shapes with randomized sizes and positions. These circular shapes are characterized by two positive pixel values and one negative pixel value. The second category of image data features three circular shapes of arbitrary sizes and random positions. However, in contrast to the first category, two of these circular shapes exhibit negative pixel values, while the remaining one displays positive pixel values (refer to Figure \ref{fig:Simulated}). The simulation results, presented in Table ~\ref{tab:table2}, compare the performance of two transport-based techniques. The first method utilizes the RSCDT transform on images followed by the nearest subspace (NS) classification technique. The second method applies the RCDT transform on the image after taking the absolute value, as RCDT can only be used on positive distributions. From the results in Table ~\ref{tab:table2}, it is evident that the RSCDT-NS method achieves 100\% accuracy in distinguishing the two classes. However, the RCDT-NS method fails to correctly classify the two groups and only achieves chance accuracy.

\begin{center}
	\begin{table}[htb]
		\centering
		\caption{\textbf{Comparative analysis of RSCDT-NS and RCDT-NS techniques on a simulated dataset}}		
		\vspace{0.15cm}
		\begin{tabular}{ c  c }
			\hline
			\hline
			Method& Accuracy (\%)\\
			\hline\\
           \vspace{0.05cm}
			RSCDT-NS & 100  \\ 
			\vspace{0.05cm}
			RCDT-NS& 49.00 \\
            \hline
            \hline
			
		\end{tabular}
  \label{tab:table2}
	\end{table}
\end{center}

\subsection{Result on 2D Geometric Shape Dataset}
\begin{figure}
    \centering
    \includegraphics{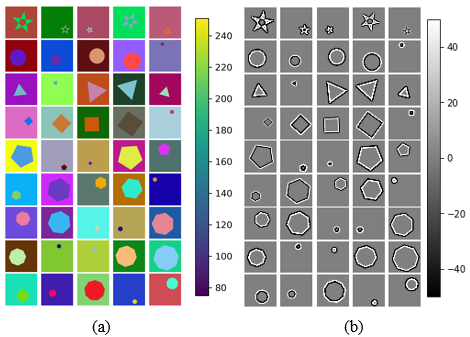}
    \caption{(a) Sample images from a 2D geometric pattern dataset. Each row in the figure represents images from a specific class. (b) Images are converted to grayscale and then Difference of Gaussian has been obtained to enhance the edge features in different shapes. This process results in the presence of negative pixel values.}
    \label{fig:geo2d}
\end{figure}

The proposed method has also been applied to a 2D geometric shape dataset \cite{Geo2D}. This dataset consists of nine geometric shapes. Each shape within a particular class can be considered as a scaled, translated, and/or rotated version of a common template, while utilizing different background and foreground colors (refer to Figure \ref{fig:geo2d}). To recognize a shape, the images have been converted to grayscale and then difference of Gaussian (DoG) \cite{DoG} has been calculated in order to enhance the edge feature. The calculation of DoG introduces negative pixel values. As the RCDT-based transport method \cite{shifat2021radon} cannot directly be implemented on images with negative pixel values, we have employed the proposed RSCDT-based nearest subspace method (RSCDT-NS) to classify the different shapes. The achieved results using RSCDT-NS have been compared with other methods, as presented in Table \ref{tab:geo2d}. It is observed that RSCDT based nearest subspace (RSCDT-NS) method outperforms all other techniques and provides an accuracy of $94.88\%$. Other transport based method RCDT-NS has also been employed directly on raw images as well as on  filtered images after taking the absolute value in order to compare with the proposed method. The performance of  both the methods are found low.   Additionally, the performances of several state-of-the-art convolutional neural network (CNN) based frameworks, say Densenet121\cite{Densenet}, ResNet-18\cite{Resnet}, ShuffleNet\cite{shufflenet} and VGG-16\cite{VGG16} have been also evaluated. In terms of classification accuracy, the best performance among the CNN-based networks has been obtained through VGG-16 using 200 training images per class  which is 79.11\%. On the other hand,  the transport-based method proposed for the signed distribution RSCDT-NS can achieve test accuracy 83.88\%  using only 20 images per class in training phase and can provide 94.88\% test accuracy using only 200 images per class in training phase.
\begin{center}
	\begin{table}[htb]
		
		\caption{\textbf{Comparative analysis of different transport based transformations and classification techniques on 2D geometric shape dataset.}}	
		\vspace{0.15cm}
		\begin{tabular}{ c  c  c  c  c  c  }
			\hline
			\hline
			\multirow{2}{*}{Method}& \multicolumn{5}{c}{No of training samples per class}  \\ 
			\vspace{0.05cm}
			& 10 & 20 & 50 & 100 &200 \\ 
			
			\hline\\

			\vspace{0.05cm}
			RSCDT-NS&  65.33& 83.88 &93.11  &92.0  &  94.88\\ 
			\vspace{0.05cm}
			RCDT-NS (Filtered image) & 51.44 &56.77  & 78.77 & 80.22 &82.55 \\ 
			\vspace{0.05cm}
			RCDT-NS (Raw image) & 12.88 &15.11  & 15.00 &16.44 &19.66 \\ 
			\vspace{0.05cm}
			DenseNet121 & 61.22 &63.44  &63.33  &63.56  &61.78 \\
   ResNet-18 &  58.83& 59.87 &64.00  &65.11  &58.33 \\
   ShuffleNet &  41.56& 44.22 &42.67  &44.42  &44.22 \\
   VGG16 &  60.44& 63.94 &69.67  &76.33  &79.11 \\
			\hline
			
			\hline
		\end{tabular}
        \label{tab:geo2d}
	\end{table}
\end{center}

\subsection{Result on sign language dataset}
Finally, we meticulously evaluate the performance of our proposed RSCDT-NS  method on a real-world sign language dataset \cite{SignL}, which encompasses images of 24 distinct classes of static signs, each representing a specific sign in sign language. Our primary objective in this study is to thoroughly assess the effectiveness of the RSCDT-NS method under various training scenarios.

\begin{figure}[htbp]
	\centering
    {\includegraphics
    [scale=0.8]{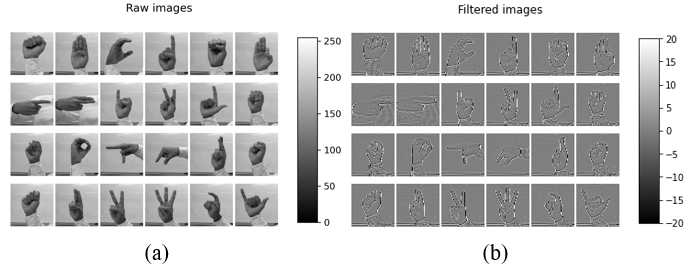}}

	\caption{(a) Images from a sign language dataset, comprising 24 types of static signs.(b) Corresponding images after undergoing difference of Gaussian (DoG) process. This operation leads to the emergence of negative pixel values. }
	\label{fig:signlanguage}
\end{figure}

To gain a deeper understanding of the dataset, we present a visualization of sample images from each class in Figure\ref{fig:signlanguage}(a). Upon closer inspection, we observe that each image in the dataset contains background pixel information, which can potentially hinder the classification task when employing available transport-based methods such as RCDT-NS. In order to mitigate the influence of the background and enhance the discriminative power of the images, we first convert the images to grayscale and subsequently calculate the difference of Gaussian(DoG) in order to enhance the edge features \cite{DoG}. This process introduces negative pixel values in the images as illustrated in Figure \ref{fig:signlanguage}(b). We then apply our proposed transport-based method, RSCDT-NS, specifically developed to handle images with negative pixel values. This preprocessing step significantly improves the suitability of the images for subsequent analysis.

Results in table \ref{tab:signlauage} show that the RSCDT-NS method outperforms the RCDT-NS method, regardless of whether the RCDT is directly applied to the raw images or the RCDT-NS is applied to the absolute filtered images. This robust performance demonstrates the superior capability of the RSCDT-NS method in effectively capturing the discriminative features inherent in the sign language dataset.

Furthermore, our comprehensive evaluation illustrates that the performance of the RSCDT-NS method surpasses that of  state-of-the-art convolutional neural network (CNN) based methods. Despite CNN based methods being a popular choice for image classification tasks, it exhibits subpar results when trained with a limited number of images per class. The best classification accuracy using CNN based technique has been achieved through ResNet-18 ($69.79\%$). In stark contrast, the RSCDT-NS method showcases its robustness and efficacy by achieving an impressive accuracy of 91.46\% using the same number of training images, firmly establishing its superiority in the field of sign language recognition.

These results provide compelling evidence that the proposed RSCDT-NS method holds significant potential for advancing signed image recognition systems, surpassing both available transport-based methods and cutting-edge deep learning approaches. By leveraging the distinctive characteristics of the sign language dataset, the RSCDT-NS method demonstrates its prowess in extracting and utilizing discriminative features, ultimately leading to highly accurate and reliable sign language recognition.

\begin{center}
	\begin{table}[htb]
		
		\caption{\textbf{Performance Evaluation of Different Transport-Based Transformations and Classification Techniques on a real sign language dataset}}	
		\vspace{0.15cm}
		\begin{tabular}{ c  c  c  c  c  c  }
			\hline
			\hline
			\multirow{2}{*}{Method}& \multicolumn{5}{c}{No of training samples per class}  \\ 
			\vspace{0.05cm}
			& 4 & 8 & 12 & 16 &20 \\ 
			
			\hline\\

            RSCDT-NS &43.75  & 65.42 & 77.71 &83.75  & 91.46 \\ 

			\vspace{0.05cm}
			RCDT-NS(Filtered image)&  40.41& 55.66 & 66.67 &76.88  &  83.96\\ 
			\vspace{0.05cm}
			RCDT-NS (Raw image) & 22.91 &30.83  &30.62  &48.33  &45.83 \\ 
			\vspace{0.05cm}
			DenseNet121 &48.75  &51.25  &49.58 &  52.92& 54.79\\
   ResNet-18 &43.33  &62.29  & 65.62 &  69.38& 69.79\\
   ShuffleNet &12.29  &30.00  & 27.92 &  47.52& 56.71\\
   VGG16 &45.83  &53.12  & 57.08 &  61.04& 60.83\\
			\hline
			
			\hline
		\end{tabular}
        \label{tab:signlauage}
	\end{table}
\end{center}

\begin{section}{Summary, Discussion, and Conclusion}

Image representation methods are important components of modern automated image analysis and classification methods. Classification problems, for example, can be difficult to solve in some representations, while dramatically simpler to solve in others. While Fourier and Wavelet transforms have been instrumental in allowing for the simple and effective solutions of numerous filtering, compression, and estimation problems, when it comes to classification, methods based on ad hoc feature extraction and deep learning have taken center stage given their simplicity and effectiveness. The lack of theoretical foundations for these methods, however, has helped slow the progress of such methodology in recent years particular in terms of performance given limited training sets, interpretability, and assurance guarantees. 

Recently methods for representing images using optimal transport have emerged \cite{aldroubi2020partitioning,kolouri2016radon,kolouri2016continuous}, and allowed for the simple non iterative (closed form) solution of a certain category of image classification problems \cite{shifat2021radon,shifat2023invariance}. Up until now, however, the methodology was limited to images that can be interpreted as probability density functions: they must be positive functions and integrate to 1. Here we have extended the previously introduced R-CDT method \cite{kolouri2016radon} by combining the R-CDT \cite{kolouri2016radon} and SCDT \cite{aldroubi2021signed} methods and introduced the Radon Signed Cumulative Distribution Transform (RSCDT). Unlike the previous RCDT method, the RSCDT now permits one to use the technique for general (signed) functions, as opposed to only positive functions that integrate to one. We described many properties of the RSCDT method (including composition and convexity of image classes, and a new mathematical distance), and described how to use these in designing simple (closed form), and accurate image classification methods using the nearest subspace technique \cite{shifat2021radon}. Comparisons with state of the art deep learning methods for image classification revealed that the newly proposed RSCDT representation method, when combined with nearest subspace methods \cite{shifat2021radon}, can yield superior accuracy. We further note that additional accuracy gains can be obtained by enhancing the nearest subspace method with additional subspace dimensions to encode affine transformations \cite{shifat2023invariance} as well as utilizing multiple ``local" subspaces to enhance the accuracy of each individual class \cite{rubaiyat2022end}. The combination of these enhanced classification methods with the newly introduced RSCDT method will be the subject of future work.

In summary, we believe that the RSCDT method introduced here will facilitate the analysis of signed functions (images) in a number of applications. By allowing for invertible transformation of signed images, the RSCDT representation method does not involve information loss. When applied to signed images that can be modeled according to the class model proposed in equation \eqref{eq:2dgenerative_model}, the transform can render image classes convex, and thus easy to partition.

\end{section}

\begin{section}{Acknowledgements}

The authors dedicate this paper to Prof. Akram Aldroubi, who has been a teacher, mentor and friend to each one of us. His expertise, wisdom, and willingness to share his knowledge have greatly contributed to our development in the field. We are truly fortunate to have had the opportunity to learn from him and to have him as a guiding force in our academic endeavors.

Authors also acknowledge funding from the NIH (GM130825) and ONR (N000142212505) for supporting this work.
\end{section}

\bibliographystyle{siam}
\bibliography{reference_merge}

\begin{thebibliography}{10}

\bibitem{aldroubi2020partitioning}
{\sc A.~Aldroubi, S.~Li, and G.~K. Rohde}, {\em Partitioning signal classes
  using transport transforms for data analysis and machine learning}, arXiv
  preprint arXiv:2008.03452,  (2020).

\bibitem{aldroubi2021signed}
{\sc A.~Aldroubi, R.~D. Martin, I.~Medri, G.~K. Rohde, and S.~Thareja}, {\em
  The signed cumulative distribution transform for 1-d signal analysis and
  classification}, Foundations of Data Science, 4 (2022), pp.~137--163.

\bibitem{Gaussianedge}
{\sc M.~Basu}, {\em Gaussian-based edge-detection methods-a survey}, IEEE
  Transactions on Systems, Man, and Cybernetics, Part C (Applications and
  Reviews), 32 (2002), pp.~252--260.

\bibitem{bracewell}
{\sc R.~N. Bracewell}, {\em The {F}ourier transform and its applications},
  McGraw-Hill Series in Electrical Engineering. Circuits and Systems,
  McGraw-Hill Book Co., New York, third~ed., 1986.

\bibitem{brown2014magnetic}
{\sc R.~W. Brown, Y.-C.~N. Cheng, E.~M. Haacke, M.~R. Thompson, and
  R.~Venkatesan}, {\em Magnetic resonance imaging: physical principles and
  sequence design}, John Wiley \& Sons, 2014.

\bibitem{cattell2017reconstructing}
{\sc L.~Cattell, C.~H. Meyer, F.~H. Epstein, and G.~K. Rohde}, {\em
  Reconstructing high-resolution cardiac mr movies from under-sampled frames},
  in Asilomar conference on signals, systems, and computers, 2017.

\bibitem{Geo2D}
{\sc A.~El~Korchi and Y.~Ghanou}, {\em 2d geometric shapes dataset--for machine
  learning and pattern recognition}, Data in Brief, 32 (2020), p.~106090.

\bibitem{DoG}
{\sc R.~C. Gonzalez and R.~E. Woods}, {\em Digital image processing, vol. 21},
  2011.

\bibitem{microscopy}
{\sc N.~Hamilton}, {\em Quantification and its applications in fluorescent
  microscopy imaging}, Traffic, 10 (2009), pp.~951--961.

\bibitem{Resnet}
{\sc K.~He, X.~Zhang, S.~Ren, and J.~Sun}, {\em Deep residual learning for
  image recognition}, in Proceedings of the IEEE conference on computer vision
  and pattern recognition, 2016, pp.~770--778.

\bibitem{Densenet}
{\sc G.~Huang, Z.~Liu, L.~Van Der~Maaten, and K.~Q. Weinberger}, {\em Densely
  connected convolutional networks}, in Proceedings of the IEEE conference on
  computer vision and pattern recognition, 2017, pp.~4700--4708.

\bibitem{pytranskit}
{\sc Imaging and data~science lab}, {\em Py\uppercase{T}rans\uppercase{K}it}.
\newblock \url{https://github.com/rohdelab/PyTransKit}.

\bibitem{scalogram}
{\sc U.~Jung and B.-H. Koh}, {\em Wavelet energy-based visualization and
  classification of high-dimensional signal for bearing fault detection},
  Knowledge and Information Systems, 44 (2015), pp.~197--215.

\bibitem{kolouri2016radon}
{\sc S.~Kolouri, S.~R. Park, and G.~K. Rohde}, {\em The {R}adon cumulative
  distribution transform and its application to image classification}, IEEE
  transactions on image processing, 25 (2016), pp.~920--934.

\bibitem{kolouri2017optimal}
{\sc S.~Kolouri, S.~R. Park, M.~Thorpe, D.~Slepcev, and G.~K. Rohde}, {\em
  Optimal mass transport: Signal processing and machine-learning applications},
  IEEE Signal Processing Magazine, 34 (2017), pp.~43--59.

\bibitem{kolouri2015}
{\sc S.~Kolouri and G.~Rohde}, {\em Transport-based single frame super
  resolution of very low resolution face images}, in Proc. IEEE CVPR, 2015,
  pp.~4876--4884.

\bibitem{kolouri2016continuous}
{\sc S.~Kolouri, A.~B. Tosun, J.~A. Ozolek, and G.~K. Rohde}, {\em A continuous
  linear optimal transport approach for pattern analysis in image datasets},
  Pattern Recognition, 51 (2016), pp.~453--462.

\bibitem{lecun1998gradient}
{\sc Y.~LeCun, L.~Bottou, Y.~Bengio, and P.~Haffner}, {\em Gradient-based
  learning applied to document recognition}, Proceedings of the IEEE, 86
  (1998), pp.~2278--2324.

\bibitem{shufflenet}
{\sc N.~Ma, X.~Zhang, H.-T. Zheng, and J.~Sun}, {\em Shufflenet v2: Practical
  guidelines for efficient cnn architecture design}, in Proceedings of the
  European conference on computer vision (ECCV), 2018, pp.~116--131.

\bibitem{natterer2001mathematics}
{\sc F.~Natterer}, {\em The mathematics of computerized tomography}, SIAM,
  2001.

\bibitem{park2017cumulative}
{\sc S.~R. Park, S.~Kolouri, S.~Kundu, and G.~K. Rohde}, {\em The cumulative
  distribution transform and linear pattern classification}, Applied and
  Computational Harmonic Analysis,  (2017).

\bibitem{SignL}
{\sc R.~F. Pinto, C.~D. Borges, A.~M. Almeida, and I.~C. Paula}, {\em Static
  hand gesture recognition based on convolutional neural networks}, Journal of
  Electrical and Computer Engineering, 2019 (2019), pp.~1--12.

\bibitem{quinto2006introduction}
{\sc E.~T. Quinto}, {\em An introduction to x-ray tomography and radon
  transforms}, in Proceedings of symposia in Applied Mathematics, vol.~63,
  2006, p.~1.

\bibitem{radon1917determination}
{\sc J.~Radon}, {\em On the determination of functions from their integrals
  along certain manifolds}, Ber. Verh, Sachs Akad Wiss., 69 (1917),
  pp.~262--277.

\bibitem{rubaiyat2020parametric}
{\sc A.~H.~M. Rubaiyat, K.~Hallam, J.~Nichols, M.~Hutchinson, S.~Li, and
  G.~Rohde}, {\em Parametric signal estimation using the cumulative
  distribution transform}, IEEE Transactions on Signal Processing,  (2020).

\bibitem{rubaiyat2022end}
{\sc A.~H.~M. Rubaiyat, S.~Li, X.~Yin, M.~S.~E. Rabbi, Y.~Zhuang, and G.~K.
  Rohde}, {\em End-to-end signal classification in signed cumulative
  distribution transform space}, arXiv preprint arXiv:2205.00348,  (2022).

\bibitem{shifat2021radon}
{\sc M.~Shifat-E-Rabbi, X.~Yin, A.~H.~M. Rubaiyat, S.~Li, S.~Kolouri,
  A.~Aldroubi, J.~M. Nichols, and G.~K. Rohde}, {\em Radon cumulative
  distribution transform subspace modeling for image classification}, Journal
  of Mathematical Imaging and Vision, 63 (2021), pp.~1185--1203.

\bibitem{shifat2023invariance}
{\sc M.~Shifat-E-Rabbi, Y.~Zhuang, S.~Li, A.~H.~M. Rubaiyat, X.~Yin, and G.~K.
  Rohde}, {\em Invariance encoding in sliced-wasserstein space for image
  classification with limited training data}, Pattern Recognition, 137 (2023),
  p.~109268.

\bibitem{VGG16}
{\sc K.~Simonyan and A.~Zisserman}, {\em Very deep convolutional networks for
  large-scale image recognition}, arXiv preprint arXiv:1409.1556,  (2014).

\bibitem{thorpeNotes}
{\sc M.~Thorpe}, {\em Introduction to optimal transport}.
\newblock \url{https://www.math.cmu.edu/~mthorpe/OTNotes}.

\bibitem{villani2003topics}
{\sc C.~Villani}, {\em Topics in optimal transportation}, no.~58, American
  Mathematical Soc., 2003.

\bibitem{villani2008optimal}
\leavevmode\vrule height 2pt depth -1.6pt width 23pt, {\em Optimal transport:
  old and new}, vol.~338, Springer Science \& Business Media, 2008.

\bibitem{wang2013linear}
{\sc W.~Wang, D.~Slepcev, S.~Basu, J.~A. Ozolek, and G.~K. Rohde}, {\em A
  linear optimal transportation framework for quantifying and visualizing
  variations in sets of images}, International journal of computer vision, 101
  (2013), pp.~254--269.

\bibitem{zhuang2022local}
{\sc Y.~Zhuang, S.~Li, X.~Yin, A.~H.~M. Rubaiyat, G.~K. Rohde, et~al.}, {\em
  Local sliced-wasserstein feature sets for illumination-invariant face
  recognition}, arXiv preprint arXiv:2202.10642,  (2022).

\end{thebibliography}

\newpage

\appendix

\begin{section}{}

\begin{subsection} {Proof that $\G$ is a group}
\label{append:GR_group}
\begin{proof}
Here we show that $(\G, \star)$ is a group. Recall, from equation \eqref{ConvexGroup},
\begin{equation*}
\G = \{g = (g^{\theta})_{\theta \in [0,\pi]} : g^{\theta} : \R \rightarrow \R \text{ is a strictly increasing bijection } \forall \theta \in [0,\pi]\}
\end{equation*}
and the operation $\star$ is defined as,
$\star : \G\times \G\rightarrow \G,$ $(g\star h) (\cdot,\theta):= (g^{\theta}\circ h^{\theta})(\cdot)$ for all $\theta \in [0,\pi)$ and  $g,h\in \G.$ Now, let's look at the individual group properties,
\begin{itemize}
    \item Closure : for any $g,h \in \G,$ $g \star h$ is component wise composition of strictly increasing functions for each $\theta \in [0,\pi],$ hence $g \star h$ is also a strictly increasing surjection. Thus, $g \star h \in \G.$
    \item Associativity : Composition of functions is always an associative operation.
    \item Identity : $I = (I^{\theta})_{\theta \in [0,\pi]}$ such that $I^{\theta}(t) = t,$ for all $ t \in \R$ and $ \theta \in [0,\pi].$ 
    \item Inverse : Note that if $\alpha$ is a strictly increasing bijection then so is $\alpha^{-1}.$ So, for any $g = (g^{\theta})_{\theta \in [0,\pi]} \in \G,$ define $g^{-1} = ((g^{\theta})^{-1})_{\theta \in [0,\pi]}$ to be the inverse of $g$. Clearly, $g \star g^{-1} = (g^{\theta} \circ (g^{\theta})^{-1})_{\theta \in [0,\pi]} = (I^{\theta})_{\theta \in [0,\pi]} = I.$
\end{itemize}
\end{proof}

\end{subsection}

\begin{subsection}{Proof of Composition Property (see Proposition \ref{composition})} \label{compo_proof}
\begin{proof}
Let
$s_g = \mathcal{R}^{-1} ((g^{\theta})' \widetilde{s} \circ g^{\theta})$ where $g \in \G$ and the composition operation is in terms of the first variable of $\widetilde s(\cdot,\theta)$.  Then by the definition of $\mathcal{R}$, we have $\widetilde{s_{g}} = (g^{\theta})' \widetilde{s} \circ g^{\theta}$, which implies the following in terms of the corresponding CDFs:  $F_{\widetilde{s}_{g}(\cdot,\theta)} = F_{\widetilde{s}(\cdot,\theta)} \circ g^{\theta}, \forall \theta$. Now the goal is to evaluate SCDT for each $\widetilde{s_g}(\cdot, \theta)$, which we denote $\widetilde{s}_{g^{\theta}}(\cdot)$ for simplicity. Similarly we denote $(\widetilde{s_g})^{\pm}(\cdot, \theta)$ as $\widetilde{s}^{\pm}_{g^{\theta}}(\cdot)$.  First note that 
 \begin{equation} \label{compo_norms}
     \|\widetilde{s}^{\pm}_{g^{\theta}}\|_1 = \|\widetilde{s}^{\pm}(\cdot, \theta)\|_1, \forall \theta.  
 \end{equation}
Indeed,  since $g^{\theta}$ is strictly increasing, we have
\begin{equation}
    \widetilde{s}^{\pm}_{g^{\theta}} = (g^{\theta})' \widetilde{s}^{\pm} \circ g^{\theta},
\end{equation}
and by the change of variables formula
\begin{align*}
    \|\widetilde{s}^{\pm}_{g^{\theta}}\|_1 = \int_{\R} \widetilde{s}^{\pm}_{g^{\theta}}(t) \, dt =  \int_{\R} (g^{\theta}(t))' \widetilde{s}^{\pm}(g^{\theta}(t),\theta )\, dt = \int_{\R} \widetilde{s}^{\pm}(u,\theta) \, du = \|\widetilde{s}^{\pm}(\cdot, \theta)\|_1. 
\end{align*}
Similarly one can show that 
\begin{equation} \label{compo_CDFparts}
    F_{\widetilde{s}^{\pm}_{g^{\theta}}} = F_{\widetilde{s}^{\pm}(\cdot, \theta)} \circ g^{\theta}, \forall \theta.
\end{equation}
Then  by definition of the SCDT in \eqref{signed},  the composition property follows form equations \eqref{compo_norms} and \eqref{compo_CDFparts} proved above. Indeed,
\begin{align*}
    \widehat{s}_{g} &= \Big((\widetilde{s}^{+}_{g^{\theta}})^{\star}, \|\widetilde{s}^{+}_{g^{\theta}}\|_1, (\widetilde{s}^{-}_{g^{\theta}})^{\star}, \|\widetilde{s}^{-}_{g^{\theta}}\|_1\Big) \\&= 
    \Big(F^{\dagger}_{\frac{\widetilde{s}^{+}_{g^{\theta}}}{\|\widetilde{s}^{+}_{g^{\theta}}\|_1}} \circ F_{\frac{\widetilde{r}}{\|\widetilde{r}\|_1}}, \|\widetilde{s}^{+}_{g^{\theta}}\|_1, F^{\dagger}_{\frac{\widetilde{s}^{-}_{g^{\theta}}}{\|\widetilde{s}^{-}_{g^{\theta}}\|_1}} \circ F_{\frac{\widetilde{r}}{\|\widetilde{r}\|_1}}, \|\widetilde{s}^{-}_{g^{\theta}}\|_1\Big)
    \\ &= \Big(\big(F_{\frac{\widetilde{s}^{+}}{\|\widetilde{s}^{+}\|_1}} \circ g^{\theta} \big)^{\dagger} \circ F_{\frac{\widetilde{r}}{\|\widetilde{r}\|_1}}, \|\widetilde{s}^{+}\|_1, \big(F_{\frac{\widetilde{s}^{-}}{\|\widetilde{s}^{-}\|_1}} \circ g^{\theta} \big)^{\dagger} \circ F_{\frac{\widetilde{r}}{\|\widetilde{r}\|_1}}, \|\widetilde{s}^{-}\|_1\Big) \\ &= 
    \Big( (g^{\theta})^{-1} \circ F^{\dagger}_{\frac{\widetilde{s}^{+}}{\|\widetilde{s}^{+}\|_1}}  \circ F_{\frac{\widetilde{r}}{\|\widetilde{r}\|_1}}, \|\widetilde{s}^{+}\|_1, (g^{\theta})^{-1} 
 \circ F^{\dagger}_{\frac{\widetilde{s}^{-}}{\|\widetilde{s}^{-}\|_1}}  \circ F_{\frac{\widetilde{r}}{\|\widetilde{r}\|_1}}, \|\widetilde{s}^{-}\|_1\Big) \\ &=  \Big( (g^{\theta})^{-1} \circ (\widetilde{s}^+)^\star, \|\widetilde{s}^{+}\|_1, (g^{\theta})^{-1} 
 \circ (\widetilde{s}^-)^\star, \|\widetilde{s}^{-}\|_1\Big), 
\end{align*}
where the second to the last equality follows from the fact that $g^{\theta}$ is strictly increasing.
\end{proof}

\end{subsection}

\begin{subsection} {Proof of the Convexification Property (see Proposition \ref{convexification})}
\begin{proof}
Assume $H_R^{-1}$ is a convex set. We show that $\widehat{S}_{\phi,H_R}$ is convex. To that end, let $\widehat{\phi}_g, \widehat{\phi}_h \in \widehat{S}_{\phi,H_R}.$ 
\begin{remark}
For the set of equations below, note that $g,h \in H_R,$ i.e. $g := (g^{\theta})_{\theta \in [0,\pi]}$ and $h := (h^{\theta})_{\theta \in [0,\pi]}.$ The equality below holds true for all $\theta \in [0,\pi]$ because the operations are done component wise. To simplify our calculations here, we avoid the cumbersome $\theta$ notation.
\end{remark}
For $\lambda \in [0,1],$ consider the convex combination,
\begin{align*}
    \lambda \widehat{\phi}_g + (1-\lambda) \widehat{\phi}_h 
    &= \lambda \Big( (\widetilde{\phi}^+_g)^\star, \|\widetilde{\phi}^+_g\|_1, (\widetilde{\phi}^-_g)^\star, \|\widetilde{\phi}^-_g\| \Big) + (1-\lambda) \Big((\widetilde{\phi}^+_h)^\star, \|\widetilde{\phi}^+_h\|_1, (\widetilde{\phi}^-_h)^\star, \|\widetilde{\phi}^-_h\| \Big) \\ &= \Big(\lambda (\widetilde{\phi}^+_g)^\star + (1-\lambda) (\widetilde{\phi}^+_h)^\star, \lambda\| \widetilde{\phi}^+_g\| + (1-\lambda) \| \widetilde{\phi}^+_h\|, \\ &\lambda (\widetilde{\phi}^-_h)^\star + (1-\lambda) (\widetilde{\phi}^-_h)^\star, \lambda\| \widetilde{\phi}^-_g\| + (1-\lambda) \| \widetilde{\phi}^-_h\| \Big)
    \\ &= \Big(\lambda (\widetilde{\phi}^+_g)^\star + (1-\lambda) (\widetilde{\phi}^+_h)^\star, \| \widetilde{\phi}^+\|, \lambda (\widetilde{\phi}^-_h)^\star + (1-\lambda) (\widetilde{\phi}^-_h)^\star, \| \widetilde{\phi}^-\| \Big)
    \\ &= \Big(\lambda (g^{-1} \circ (\widetilde{\phi}^+)^\star) + (1-\lambda)(h^{-1} \circ (\widetilde{\phi}^+)^\star), \| \widetilde{\phi}^+\|, \\ & \lambda (g^{-1} \circ (\widetilde{\phi}^-)^\star) + (1-\lambda)(h^{-1} \circ (\widetilde{\phi}^-)^\star), \| \widetilde{\phi}^-\| \Big) 
    \\ &= \Big( (\lambda g^{-1}  + (1-\lambda)h^{-1}) \circ (\widetilde{\phi}^+)^\star, \| \widetilde{\phi}^+\|, (\lambda g^{-1}  + (1-\lambda)h^{-1}) \circ (\widetilde{\phi}^-)^\star, \| \widetilde{\phi}^-\| \Big) 
    \\ &= \Big( p^{-1} \circ (\widetilde{\phi}^+)^\star, \| \widetilde{\phi}^+\|, p^{-1} \circ (\widetilde{\phi}^-)^\star, \| \widetilde{\phi}^-\| \Big) 
\end{align*}
 where $p^{-1} =  \lambda g^{-1}  + (1-\lambda)h^{-1} \in H_R$ and the norms $\|(\widetilde{\phi})^{\pm}_g\|_1 = \|(\widetilde{\phi})^{\pm}_h\|_1 = \|(\widetilde{\phi})^{\pm}\|_1,$ as shown in the proof of Composition property \ref{compo_proof} above. Thus, $\lambda \widehat{\phi}_g + (1-\lambda) \widehat{\phi}_h  \in \widehat{S}_{\phi,H_R}.$

 \end{proof}
\end{subsection}

\begin{subsection}{Proof that $D(\cdot,\cdot)$ is a metric (see def. \ref{signed_sliced_wasser}) and $D$ induces and isometry (see prop. \ref{isometry_rscdt})}
\begin{proof}
Here we show that $D$ is a metric. Recall by definition 
\ref{signed_sliced_wasser},

\begin{align*}
    D(s_1,s_2) :&= \Bigg(\int_{0}^{\pi}  
    D_{S}^2 \left({\widetilde{s_{1}}}(\cdot,\theta),{\widetilde{s_{2}}}(\cdot,\theta)\right) d\theta \Bigg)^{\frac{1}{2}}
    \\ &= \Bigg( \int_{0}^{\pi}  
    D_{W^2}^2 \left({\widetilde{s_{1}}}^+(\cdot,\theta),{\widetilde{s_{2}}}^+(\cdot,\theta)\right) d\theta + \int_{0}^{\pi} D_{W^2}^2 \left({\widetilde{s_{1}}}^-(\cdot,\theta),{\widetilde{s_{2}}}^-(\cdot,\theta)\right)
     d\theta \Bigg)^{\frac{1}{2}},
\end{align*}
where $D_S$ and $D_{W^2}$ are as defined in \eqref{SCDT_Metric}.

\begin{itemize}
    \item Non - Negativity ($D(\cdot,\cdot) \geq 0$) : Note that  $D_S^2(\cdot, \cdot) \geq 0,$ therefore, $D(\cdot,\cdot) \geq 0.$
    \item $D(s_1, s_2) = 0 \implies s_1 = s_2 : $ Assume,
        $D(s_1, s_2) = 0 .$ This implies, $ \int_{0}^{\pi}  
    D_{S}^2 \left({\widetilde{s_{1}}}(\cdot,\theta),{\widetilde{s_{2}}}(\cdot,\theta)\right) d\theta  = 0.$ Since, $D_S^2(\cdot,\cdot) \geq 0,$ therefore, $ D_{S} \left({\widetilde{s_{1}}}(\cdot,\theta),{\widetilde{s_{2}}}(\cdot,\theta)\right) = 0 \quad \forall \theta \in [0,\pi].$ Hence, $ {\widetilde{s_{1}}}(\cdot,\theta) = {\widetilde{s_{2}}}(\cdot,\theta) \quad \forall \theta \in [0,\pi]$ which further implies, $s_1 = s_2.$
    \item Symmetry ($D(s_1, s_2) = D(s_2, s_1)$) : This follows from the fact that $D_S(\cdot,\cdot)$ is a metric (see \cite{aldroubi2021signed}) and hence symmetric. Therefore, $D_S^2({\widetilde{s_{1}}}(\cdot,\theta),{\widetilde{s_{2}}}(\cdot,\theta)) = D_S^2({\widetilde{s_{2}}}(\cdot,\theta),{\widetilde{s_{1}}}(\cdot,\theta)) \quad \forall \theta \in [0,\pi].$ Integrating both sides with respect to $\theta \in [0,\pi],$ we have $D(s_1, s_2) = D(s_2, s_1).$
    \item Triangle inequality ($D(s_1, s_2) \leq D(s_1, r) + D(r, s_2)$) : By definition,
    \begin{align*}
        D^2(s_1,r) &= \int_{0}^{\pi} D^2_S(\widetilde{s_1}(\cdot,\theta), \widetilde{r}(\cdot,\theta)) \, d\theta \\
        &= \int_{0}^{\pi} \| \widehat{s_1}(\cdot,\theta) - \widehat{r}(\cdot,\theta)\|^2_{L^2(\widetilde{r}(\cdot,\theta))} \, d\theta \\
        &= \int_{0}^{\pi} \int_{\R} (\widehat{s_1}(t,\theta) - t)^2 \widetilde{r}(t,\theta) \,dt \, d\theta
    \end{align*}
    where the second equality follows from the isometry of $D_S(\cdot,\cdot)$ \cite{aldroubi2021signed}. Similarily, 
    \begin{equation}
        D^2(s_2,r)  = \int_{0}^{\pi} \int_{\R} (\widehat{s_2}(t,\theta) - t)^2 \widetilde{r}(t,\theta) \,dt \, d\theta
    \end{equation}
    Let $h(t,\theta)$ be the RSCDT of $s_2$ with respect to $s_1,$ i.e.
    \begin{equation} \label{dist}
        D^2(s_2,s_1) = \int_{0}^{\pi} \int_{\R} (h(t,\theta) - t)^2 \widetilde{s_1}(t,\theta) \,dt \, d\theta
    \end{equation}
    Following the composition property, we have $h(\cdot,\theta) \circ \widehat{s_1}(\cdot,\theta) = \widehat{s_2}(\cdot,\theta)$  and $\widetilde{r}(\cdot,\theta) = \widehat{s_1}'(\cdot,\theta) \widetilde{s_1}(\widehat{s_1}(\cdot,\theta),\theta)$ for every $\theta \in [0,\pi].$ Using change of variables $t = \widehat{s}(u,\theta)$ in \eqref{dist},
    \begin{align*}
        D^2(s_2,s_1) &= \int_{0}^{\pi} \int_{\R} (h(\widehat{s_1}(u,\theta),\theta) - \widehat{s_1}(u,\theta))^2 \widetilde{s_1}(\widehat{s_1}(u,\theta),\theta) \, \frac{d\widehat{s_1}(u,\theta)}{du} du \, d\theta \\
        &= \int_{0}^{\pi} \int_{\R} (\widehat{s_2}(u, \theta) - \widehat{s_1}(u,\theta))^2 \widetilde{r}(u,\theta) \, du \, d\theta \\
        &= \| \widehat{s_2} - \widehat{s_1} \|^2_{(L^2(\widetilde{r}) \times L^2[0,\pi])^2} \\
    \end{align*}
    This establishes the isometry i.e. $D(s_2,s_1) = \| \widehat{s_2} - \widehat{s_1} \|_{(L^2(\widetilde{r}) \times  L^2[0,\pi])^2}.$ Using this relation, we finally have the required triangle inequality for the metric,
    \begin{align*}
        D(s_2,s_1) &= \| \widehat{s_2} - \widehat{s_1} \|_{(L^2(\widetilde{r}) \times L^2[0,\pi])^2} \\ 
        &\leq \| \widehat{s_2} - I \|_{(L^2(\widetilde{r}) \times L^2[0,\pi])^2} + \| I - \widehat{s_1} \|_{(L^2(\widetilde{r}) \times L^2[0,\pi])^2} \\
        &= D(s_2,r) + D(r,s_1)
    \end{align*}

\end{itemize} 

\end{proof}
\end{subsection}

\end{section}

\end{document}